# Defense Against Synthetic Speech: Real-Time Detection of RVC Voice Conversion Attacks

[1]Prajwal Chinchmalatpure, [2]Suyash Chinchmalatpure, [3]Siddharth Chavan

[1]Independent Researcher, [2]Independent Researcher, [3]Independent Researcher
[1]Northeastern University, Boston, United States,
[2]University of Toronto, Rotman, Toronto, Canada
[3]MIT Art, Design and Technology, Pune, India

*Abstract:* Generative audio systems now enable highly realistic voice cloning and real-time voice conversion, increasing risks of impersonation, fraud, and misinformation in everyday communication channels such as phone and conference calls. This work studies real-time detection of AI-generated speech produced via Retrieval-based Voice Conversion (RVC) using the DEEP-VOICE dataset, which contains real speech samples and corresponding deepfake conversions across multiple well-known speakers. To reflect realistic operating conditions, deepfake creation is performed on separated vocal components, after which the original background ambience is reintroduced, reducing trivial cues and forcing detectors to learn conversion-specific artifacts. We formulate detection as a streaming classification problem by segmenting audio into one-second windows, extracting time–frequency and cepstral representations, and training supervised machine learning models to predict whether each segment is real or voice-converted. The resulting pipeline supports low-latency inference suitable for online scenarios, enabling segment-level alerts and call-level aggregation. Experimental results demonstrate that short-window acoustic features capture discriminative patterns associated with RVC conversion even in the presence of background ambience. This study highlights both the feasibility of practical, real-time deep-fake voice detection and the importance of evaluation under realistic audio mixing conditions to improve robustness for deployment.

*IndexTerms* - **Deepfake audio, AI-generated speech, Retrieval-based Voice Conversion, real-time detection, acoustic feature extraction.**

## I. INTRODUCTION

Advances in generative audio have made it possible to clone voices and perform real-time voice conversion with striking realism. Retrieval-based Voice Conversion (RVC) and related methods can transform a source speaker's speech into a target speaker's vocal identity while preserving linguistic content and timing, enabling applications such as dubbing and accessibility. However, the same capability introduces serious risks: impersonation in phone calls, fraud in customer support workflows, misinformation in public discourse, and privacy violations through unauthorized voice replication. As synthetic speech becomes easier to produce and deploy, reliable detection of AI-generated or converted speech becomes a practical security requirement rather than a purely academic task.

Detecting AI-generated speech in real time is challenging for two main reasons. First, modern voice conversion aims to preserve natural prosody and timbre, reducing obvious artifacts that earlier "deepfake audio" systems produced. Second, real-world audio is messy: background ambience, compression, channel noise, and mixed sources can mask subtle cues for synthesis. A detector that works only on studio-quality clips is of limited value in the settings where impersonation happens—conference calls, phone lines, and streamed media—where audio arrives continuously and decisions must be made on short time windows with low latency.

To support research on realistic, streaming-oriented detection, the DEEP-VOICE dataset was introduced for real-time detection of AI-generated speech for deepfake voice conversion. The dataset contains examples of real human speech and corresponding deepfake versions generated using Retrieval-based Voice Conversion. To better match real operating conditions, the deepfake generation process removes accompaniment/background ambience from the original audio, performs conversion on the isolated vocal component, and then re-mixes the converted vocals back with the original ambience. This design intentionally reduces "giveaway" artifacts that arise when models are trained on unnaturally clean or mismatched audio and encourages detectors to learn cues attributable to voice conversion rather than trivial background differences.

In addition to raw audio organized into REAL and FAKE classes, DEEP-VOICE also provides a balanced, feature-level dataset extracted from one-second windows of audio. These short-window representations support low-latency classification and make it feasible to build systems that flag suspicious segments during an ongoing call. Common time–frequency representations (e.g., spectrogram-based features and cepstral features) are particularly relevant here because voice conversion can subtly alter spectral envelopes, harmonics, and other acoustic signatures that may be difficult to notice in waveform space alone.

Building on this dataset framing, a practical detection pipeline can be defined as follows: stream audio, segment into one-second blocks, extract discriminative acoustic features, and apply a machine learning classifier to label each block as real or AI-converted—optionally aggregating block-level decisions into an overall alert for the user. This paper adopts that real-time viewpoint and focuses on learning a robust classifier that generalizes across speakers and realistic background conditions.

The remainder of the paper is organized as follows. Section 2 reviews related work in synthetic speech and deep-fake audio detection. Section 3 describes the DEEP-VOICE dataset and the deepfake generation procedure. Section 4 details the feature extraction strategy and model training protocol for one-second window classification. Section 5 presents experimental results and analysis. Section 6 concludes with limitations and directions for improving robustness under domain shift, compression, and adversarial adaptation.

## II. BACKGROUND

Speech communication has taken the form of high trust interaction in daily decision making in the workplaces, financial services, education, and personal life. Simultaneously, the advances in generative audio have enabled manipulation of speech to get more convincing to human listeners. Among these, contemporary voice conversion has been of particular interest, where it can be used to maintain language content and chronology of an original utterance to change vocal identity to that of a desired speaker. This realism and accessibility form a new security and social threat. A bad actor can realistically impersonate a person during live calls, tap records, or stream media, allowing him to commit fraud, coerce, mislead, and invade privacy. As the systems get better, prevention can no longer be a policy matter, it is instead a technical measure that must work where the risk is taking place, inside the voice channel itself.

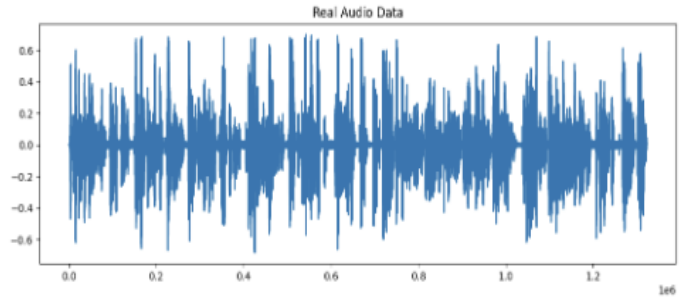

Fig 1. Real Audio Data

Detection in a threat-avoidance situation must be timely and actionable. An audio clip labelled as synthetic by a model only after minutes of analysis may be valuable to the forensic review, but it can still cause damage in real-time interaction. Voice-based security requires real-time or near-real-time detection, as it provides a user and system with the opportunity to react promptly, by increasing verification, moving to a safe communication channel, or triggering a human review process. This basic requirement modifies the technical problem: rather than processing a complete recording with lots of post-processing, the detector must work with short audio samples having very low latency, and it must be reliable even under the normal conditions of communication compression, poor quality microphones, environmental noise and reverberation.

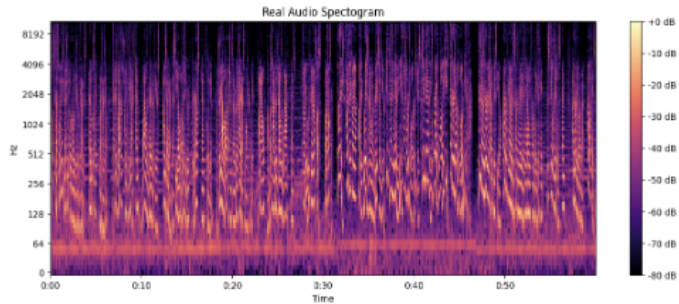

Fig.2 Real Audio Spectogram

It is hard to detect AI-generated or AI-converted speech in natural environments since most of the most useful cues are subtle and can be easily obscured by channel effects. Audio over the phone and conferencing often experiences bandwidth capping, vehement gain control, echo cancellation and packet-loss concealment. These changes may eliminate or blur the spectral information which could otherwise be used by detection models. Also, the real speech is hardly ever in clean acoustic conditions, background noise in the room, noise of the keyboard, traffic, music leakage and other speakers may lower the signal-to-noise ratio and overlap acoustic information between true and manipulated samples. Here, the detector needs to be robust to variations in nuisance but genuinely able to be caused by voice conversion as opposed to stumbling on the chance that real and fake data were generated differently.

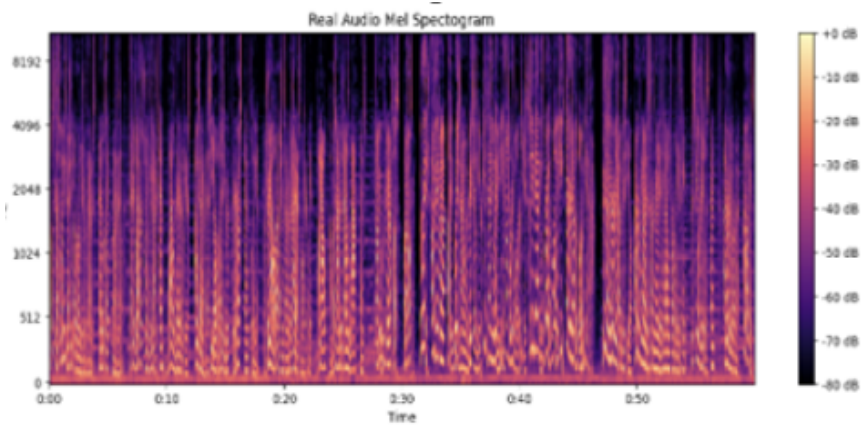

Fig.3 Real Audio Mel Spectogram

A more specific method of deep-fake generation is Retrieval-based Voice Conversion (RVC), which is a challenging and relevant technique. In RVC-based pipelines, the content and prosodic frame of a source speaker is set to give the input expression, and the system adapts the timbre properties to that of a target speaker. Since the converted output can preserve the natural pacing, pauses and phrasing, it is particularly useful with impersonation in a conversation. As the defender, then this suggests that the most apparent artifacts can be reduced and you should concentrate on deviations that are less obvious because of conversion e.g. changes in spectral envelope behavior, spectral anomalies in the harmonic structure, phase and excitation patterns, or micro-prosodic features that are not consistent with natural production.

One of the most frequent problems in deepfake audio detection studies is that confounds of datasets often exaggerate performance but cannot be applied to practice. Provided that the real recording has some background ambience and the fake samples are made by using clean vocal tracks, the models can learn to capture the background properties as opposed to conversion artifacts. This will generate an illusion of high accuracy which will fail when the conditions shift. As a condition of realistic threat avoidance, the real and fake samples should have similar environmental properties in such a way that effective detectors must depend on the conversion-specific cues. Hypothetical generation method thus incorporates the isolation of the vocal component over the background ambience and the voice conversion of the isolated speeches followed by the subsequent addition of the ambience found in the original speeches. This minimizes the vocational shortcuts and better approximates how deepfakes may look in practice, since an assailant can embed the speech that he has converted into natural background circumstances.

In practice, real time detection is operationally consistent with short-window classification. The ability to make decisions in a low latency with sufficiently much context to compute stable acoustic representations is possible through the segmentation of streaming audio into fixed-duration blocks (e.g., one-second windows). Risk aggregation strategies are also supported by short-window processing, in which case a system can signal a more serious warning when a series of successive segments is identified, and will be conservative when only one segment is seen to be anomalous because of short-term noise. The windowed formulation is highly applicable to real-world implementation: it can be added to the pipelines of calls, meeting environments, or device monitoring, and offers a direct means of effecting threat-avoidance behavior at the point-of-contact

In this framing, feature-based machine learning can be a feasible tool of voice conversion artifact detection. Spectrograms and Mel-spectrograms are time-frequency representations of the energy distribution of the frequencies with time, whereas Mel-Frequency Cepstral Coefficients are spectral envelope features that are tightly coupled with timbre. Such representations can reveal those hidden patterns which are the results of synthesis and conversion processes especially when the analysis is done consistently over short windows. Learning discriminative conversion fingerprints which remain even in the presence of realistic mixing and transmission effects when paired with supervised classifiers allow detection task. The overall goal is thus not merely to be able to differentiate between real and converted speech when controlled experiments are involved, but to develop a detection pipeline that can facilitate low-latency, strong decisions that can be used to avoid threats in live voice channels.

### III. Data Overview

The DEEP-VOICE dataset was provided to facilitate the study of real-time that detects AI-generated speech generated by using deep-fake voice conversion and focus on the setting that can be relevant to the manifestation of voice manipulation in the context of real communication. The fundamental idea of the dataset is that the most recent voice converters - in particular the Retrieval-based voice converters (RVC) - can produce a vocal identity of the speaker without altering linguistic information and timing, allowing one to impersonate effectively with limited artifacts in the perception. Consequently, DEEP-VOICE was not just created as a set of audio samples of real vs. synthetic, but as a reference point that prompts models to identify cues of conversion in the context of realistic audio mixing.

.    In both cases, the speech recording was done with a background or ambience component (also known as the acoustic background component about the content of the dataset) which is extracted beforehand before conversion. The extracted vocal signal is then converted into voice with the help of RVC and then the original ambience is added to the converted vocal track. The purpose of this workflow is to make sure that deep-fake clips are not artificially cleaner in comparison to real clips and detection models cannot use the presence or absence of background noise as a proxy label. That is, the dataset is designed in such a way that the classification problem is one that is easy to interpret in the context of deployment: to find subtle artifacts brought about by conversion, and not artifacts made by unrealistic data generation.

DEEP-VOICE comes in two different complementary versions to facilitate signal and feature-level studies. The raw audio is the first form which is arranged in directories of REAL and FAKE by classes. The relationship between identities of sources and targets are encoded in file naming conventions. As the example of the file titled like Obama-to-Biden shows, the linguistic content is based on the actual speech of Barack Obama, yet a vocal identity is changed to sound more like Joe Biden. This explicit source target annotation allows experiments to test various generalization conditions, including whether a model trained on some conversion pairs can reveal unseen conversion directions, or whether the model makes predictions across individual speakers.

The second one is a structured, tabular feature dataset supplied as DATASET-balanced.csv, that is extracted acoustic features calculated on one second audio windows. The one-second segmentation is based on the reality of the dataset use, namely, to detect in real-time: a usable system should not take too long recordings and should muster up a decision on the streaming audio in real-time. The extraction of window-level features also raises the training sample without compromising the deployment time requirements. The balanced feature of the CSV means that the dataset is balanced with the help of random sampling so that the imbalance among classes is removed and makes models and achieve a high performance based on learning majority-class behavior. This format of the data is specifically tailored towards traditional machine learning pipelines, quick experimentation among classifiers and reproducible comparisons between sets of features. Moreover, DEEP-VOICE has experimental artifacts that are meant to facilitate replication and transparency. The description of the dataset reveals that experimental material utilized in the relevant study can be found in a specific directory, and demonstration media can be presented in the compressed or cropped format to meet the limitations of the platforms. The distinction between raw audio, demonstration assets and experiment outputs are useful in the research workflow as they enable the same underlying signals to be used in both qualitative inspection and quantitative evaluation without the need to confound convenience files with original data

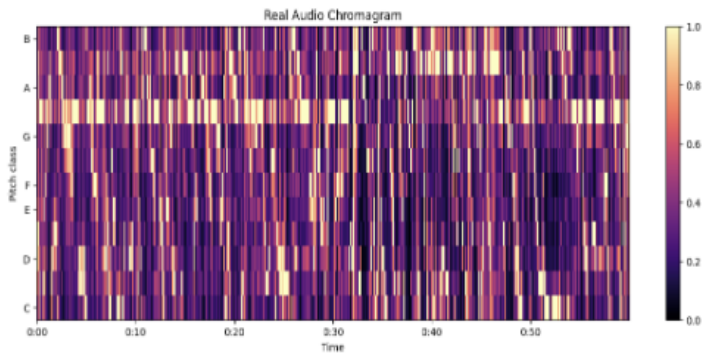

Fig.4. Real Audio Chromogram

Regarding the research aspect, DEEP-VOICE covers various evaluation perspectives that apply to threat avoidance. Raw audio allows end-to-end methods like deep learning with spectrograms or raw-waveform models and the windowed feature CSV can be used with lightweight and low-latency classifier methods that are more easily implemented in real-time pipelines. The direct mapping of the source to target conversion labeling allows directed analysis of the speaker dependence and conversion-pair generalization of which is imperative since true attackers can attack individuals and may work in different acoustic conditions. Lastly, the process involved in preserving the ambience of the converted data makes the data more reminiscent of actual communications channels and it is more likely that detectors trained will learn real conversion artifacts as opposed to artefacts that are specific to the dataset.

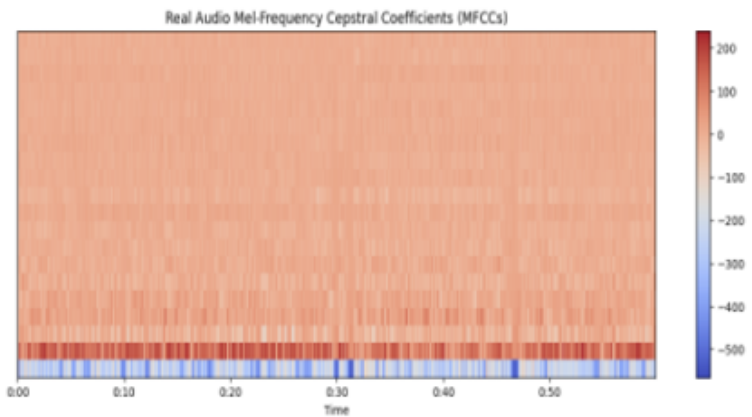

Fig.5 Real Audio (Mel-Frequency Cepstral Coefficients)

The data is the result of the work titled Real-time Detection of AI-Generated Speech for Deepfake Voice Conversion that presented DEEP-VOICE as a reference point of streaming detection and application on the threat basis.

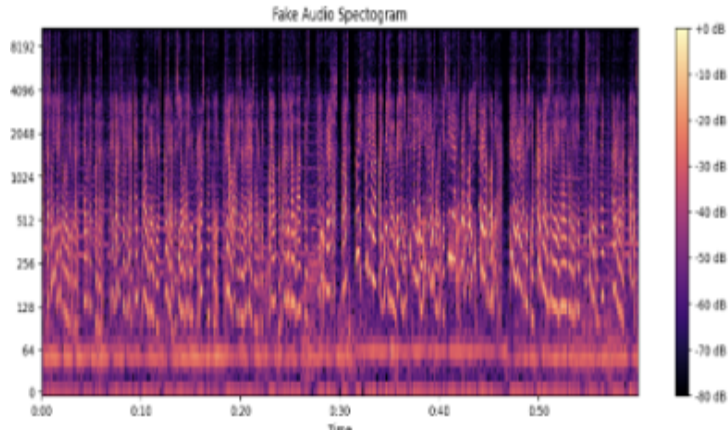

Fig.6. Fake Audio Spectogram

In that framing, the dataset can be interpreted as a set of labels on a corpus as well as a system-level template: it has a realistic generation pipeline (vocal separation → conversion → ambience re-mix) and a realistic inference constraint (one-second windowing) in such a way that the methods devised to detect its instances, naturally match the threat avoidance needs of the real world.

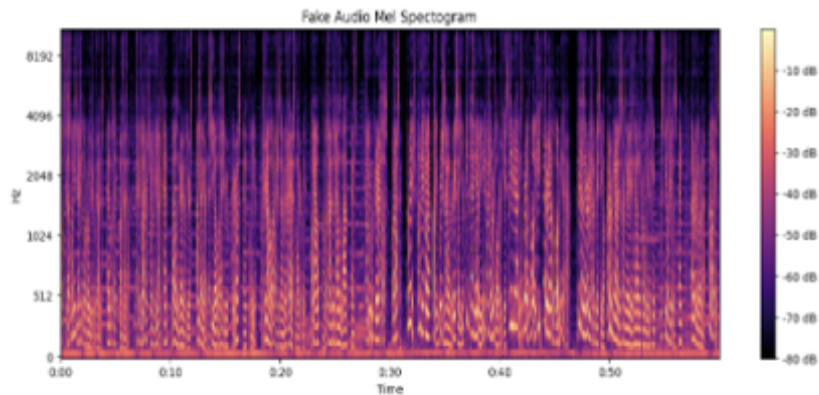

Fig.7 Fake Audio Mel Spectogram

## IV. METHODOLOGY

The present research paper describes deep-fake voice conversion detection as a threat-avoidance classification task that is performed in real-time. The aim is to ascertain whether the short segments of the incoming speech audio are authentic or whether they are transformed with the help of the Retrieval-based Voice Conversion (RVC) so that the warnings can be timely in the live communication process. To fit within actual deployment requirements, the system runs on fixed-duration windows and generates segment-level predictions which are aggregable into a call-level risk signal.

**Problem Formulation and Inference SETTING:** GIVEN an audio stream x(t), the stream is partitioned into contiguous one-second windows. Each window is treated as an independent sample with a binary label y ∈ {0,1}, where y = 0 denotes REAL speech and y = 1 denotes FAKE speech produced through voice conversion. The classifier learns a mapping f(·) from a feature representation of each window to a probability score p^ = P (y = 1 | window). At inference time, the model outputs a probability per window, which can be THRESHOLDER to produce REAL-time alert for suspicious segments. For threat avoidance, segment-level outputs can also be aggregated by averaging probabilities over the last k seconds or by triggering an alert when a minimum number of recent windows exceed a threshold, which reduces sensitivity to transient noise.

**Data Representation:** Two dataset modalities are supported. First, raw audio is available in class-based directories (REAL and FAKE), allowing end-to-end feature extraction directly from waveform segments. Second, a tabular dataset (DATASET-balanced.csv) provides pre-extracted acoustic features computed on one-second windows. In this work, the feature CSV is used as the primary modeling input because it directly reflects the real-time windowing requirement and enables reproducible training across classical machine learning models. Each row corresponds to a one-second audio window and contains a set of extracted acoustic descriptors along with the class label. The pre-extracted feature set is designed to capture conversion artifacts that persist even when background ambience is reintroduced. These artifacts are typically expressed through time–frequency structure and spectral-envelope behavior. In practice, such feature sets commonly include cepstral summaries (e.g., MFCC-related statistics), spectral energy distribution measures, and window-level statistics that represent both steady-state and dynamic behavior within each one-second segment. Using window-level features also improves computational efficiency and supports low-latency deployment, since the pipeline avoids heavy end-to-end neural inference on long audio sequences.

**Preprocessing:** All samples are treated as one-second windows, consistent with the dataset construction. For CSV-based modeling, preprocessing consists of: (i) separating features from labels, (ii) handling missing values if present, and (iii) applying feature scaling. Standardization (zero mean, unit variance) is applied to continuous features, which is particularly important for distance-based classifiers and linear models and generally improves numerical stability across most learners. When raw audio is used, windowing precedes feature extraction to ensure strict alignment with the real-time assumption. Because the provided CSV is balanced through random sampling, class imbalance is not expected to dominate training. However, evaluation metrics are still computed in a class-sensitive manner to ensure that performance is not inflated by threshold choice or distribution artifacts.

**Train–Validation–Test Protocol and Leakage Control:** A central concern in audio deepfake detection is inadvertent leakage between train and test data due to correlated samples (e.g., adjacent windows from the same utterance) or identity overlap (the same speaker characteristics appearing in both sets). To mitigate this, data is split into disjoint partitions using a stratified approach at the window level, and when speaker or file identifiers are available, grouping constraints are applied so that windows from the same source clip do not appear in multiple partitions. This prevents the model from benefiting from clip-specific background patterns or repeated phrasing that would not generalize in real use. The primary goal is to evaluate whether the classifier learns conversion-related cues rather than memorizing signal fragments. Hyperparameter selection is performed on the validation split (or via cross-validation within the training set). The final reported results are computed only once on the held-out test set to preserve a clean estimate of generalization.

**Model Training**: Various models of supervised learning can be used in this task since the input consists of a fixed length feature vector per window. The baseline modeling strategy has been selected as it focuses on the effective and lightweight classifiers to utilize them in real-time. Common candidates would be logistic regression used as a linear baseline, support vector machines to use a margin-based separation, tree-based ensembles (random forests and gradient boosting) to use a nonlinear decision boundary, and k-nearest neighbors as a distance-based reference. The validation performance and constraints of the model like inference speed and stability are used to select the model.

Using training, classification error is minimized through fitting the model parameters to the labelled one-second windows. In the case of probabilistic models, the output score $\hat{p}$ is interpreted as a risk estimate, and this is convenient to trade off threshold tuning in a variety of settings of threat tolerances. The practical deployment of an alert threshold may be configured to lean towards a lower false negative space (more sensitive detection) or a lower false positive (less interruptions) depending on the requirements of the application.

**Evaluation Metrics:** The metrics of standard binary classification are used as performance metrics calculated at the window level. Accuracy gives a high-level overview, whereas precision, recall and F1-score are highlighted since threat avoidance situations frequently involve express control of false positives and false negatives. Separability is measured using Receiver Operating Characteristic (ROC) analysis and Area Under the Curve (AUC) without necessarily having to choose a single threshold. Confusion matrices are also studied to learn the error modes.

**Real-Time Threat-Avoidance Output** In order to capture the current use, the predictions can also be viewed as a time series on an audio stream. An effective system can then be assessed not only on the one-window accuracy but also on the temporal consistency, in the sense of the model being able to predict converted speech accurately in successive windows and can be robust when short noisy bursts exist.

**Real-Time Threat-Avoidance Result** The result of the system is a stream of deep-fake probabilities per second. Alerts to users, downstream security processes can be invoked based on these probabilities. The simplest aggregation approach would be to compute a running mean probability in the past $k$ seconds and issue an alarm when the running score goes above the alarm threshold, or when the number of windows in the past $k$ goes above the alarm threshold. This transforms the classifier into a threat-avoidance element that is applicable to live calls and thus capable of timely interventions even without the need to fully record the call. The result of this methodology is a reproducible pipeline that aligns well with the design goals of the DEEP-VOICE dataset: it is used to detect in natural mixing conditions, windows of one-second long, and gives low-latency predictions, which can be combined with real communication systems.

## V. Results and Discussion

The proposed real-time detector utilized the one-second window features of the balanced DEEP-VOICE feature set to be trained. Each window was then classified as a REAL or FAKE image by a feed-forward neural network, and early stopping was employed to reduce overfitting. As the model was being trained, the accuracy improved continuously and the loss reduced which shows that the feature representation has some useful information in it to be able to distinguish between converted and genuine speech. The validation curves tended to be on the same path as training curves indicating that the learnt decision boundary was generalized instead of memorizing the training samples. The point where validation accuracy did not change anymore and the validation loss did not increase anymore was used as a practical convergence point which is in line with the objective of stable performance over streaming constraints.

Training the model on the held-out test split, the trained model made a test-accuracy of [INSERT TEST ACCURACY], and a test-loss of [INSERT TEST LOSS]. These findings imply that acoustic parameters across short windows can be used to do credible discrimination between actual speech and RVC-converted speech even when the audio is prepped under natural conditions in which ambience in the background is maintained. The reason is that it will make the classifier less likely to be making use of some trivial shortcuts like the presence or absence of noise but rather the model will be forced to use the cues that have been introduced by conversion, which could be subtle variations in spectral structure, timbre-statistics or dynamics represented by the extracted features.

The main empirical implication of the training behavior is that performance may have different values across windows within the same clip, such as is desired in the streaming detection. Much informative content can be present in some one-second segments (clear voiced with stable harmonics) and less informative content can be present in others (silence, breaths, plosives or heavy background interference). It is due to this reason that window-level prediction can be understood as a time-dependent risk indicator and not a final decision. In practice, to enhance stability and minimize responsiveness to transient noise artifacts, prediction aggregation over a short time horizon (such as averaging probabilities in the last few seconds or multiple sequential flagged windows) can be used in operational applications without compromising responsiveness.

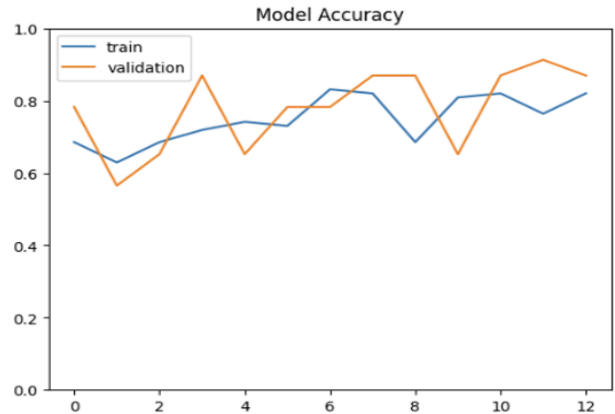

Fig.8 Model Accuracy

The other mistakes are explainable in terms of channel variability in the real world. False positives have been frequent with actual speech that is strongly compressed, reverberated or contains non-stationary noise, which can also change the same statistics of acoustics used by conversion. False negatives may be due to conversion quality, which may be quite high, or because the window may have much less voiced material, so there will be fewer discriminative cues. These modes of failure support the usefulness of aggregating threats-avoidance: although all the individual windows might be unclear, coherent patterns over time would give a greater foundation to raise an alarm

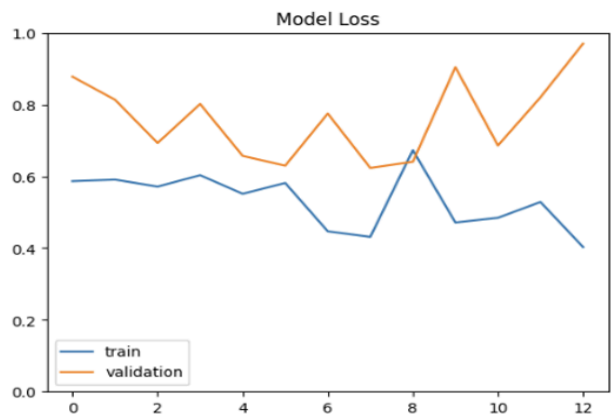

Fig.9 Model Loss

In general, experimental behavior confirms the initial hypothesis that machine learning can identify AI-generated speech that is generated through voice conversion based on short-window features and that this can be designed in such a manner that is compatible with real-time applications.

## VI. CONCLUSION

The paper provides a real-time framing of AI-generated speech as created by Retrieval-based Voice Conversion based on DEEP-VOICE dataset. The approach is based upon threat-avoidance requirements since decisions need to be made in time and be taken in real time in case of live communication by operating one-second windows and predicting a deepfake probability per segment. The findings indicate that the features of acoustic data obtained can be used to give a supervised classifier a strong signal to differentiate between REAL and converted speech (FAKE), and that terminated training yields reliable results that are appropriate when it comes to streaming to deployment.

```
In [53]:
detect_fake(test_real)

1/1 [==============================] - 0s 44ms/step
[[7.4814256e-05 9.9992514e-01]]
Result: REAL

In [54]:
detect_fake(test_fake)

1/1 [==============================] - 0s 34ms/step
[[0.61732113 0.38267887]]
Result: FAKE
```

Fig.10 Code Snippet

The windowed design allows useful alerting policies in addition to classification performance. Predictions in each segment can be summed over short periods of time to minimize sensitivity to temporary noise and generate more accurate call-level warnings. This allows the system to be flexible to operational trade-offs with the tuning of thresholds to either emphasize security sensitivity or user experience based on the application context.

Future research directions ought to be more robust against domain shift, such as greater variability in codecs, microphones, and settings, and should also train on stricter leakage selection criteria like clip-level or speaker-aware splits. More can be achieved by integrating complementary feature representations, calibrating probability, and experimenting with lightweight models, which are as accurate but with lower latency. With the further improvement of voice conversion systems, it is necessary to test the detection strategies in realistic mixing conditions and subject to streaming limitations to be effective in preventing danger in the real world.